\documentclass[epj,twocolumn]{webofc}
\usepackage[varg]{txfonts}   
\woctitle{Hadron Collider Physics symposium 2012}
\begin{document}
\title{BSM constraints from EW measurements}

\author{Celine Degrande\inst{1}\fnsep\thanks{\email{cdegrand@illinois.edu}} 
}

\institute{Department of Physics, University of Illinois at Urbana-Champaign\\
1110 W. Green Street, Urbana, IL 61801, USA          }

\abstract{%
 We investigate the impact of heavy new physics on W bosons productions at hadron colliders using an effective field theory. After listing all the relevant dimension-six operators, their effects are computed taking into account LEP constraints. Additionally, the effective approach is compared to anomalous couplings regarding several issues like unitarity. 
}

\renewcommand*\arraystretch{1.2}
\maketitle
\section{Introduction}

There are many reasons to expect new physics beyond the Standard Model (SM). The Higgs boson mass corrections depend quadratically on the cut-off of the theory and claim for new physics near the weak scale. The absence of a dark matter candidate or the lack of an explanation for the dark energy are other examples. 
If the SM is unlikely the final theory, its good agreement with the experimental data suggests that it is nevertheless a good approximation at the energies that have been probed so far. Consequently, any new model should reduce to the SM in this low energy limit. 
Nonetheless, the amount of BSM models satisfying this requirement is large and encourages model independent searches. Additionally, they allow to quantify the space left for new physics if no deviation from the SM is found. Model independent searches are based on simple assumptions valid for large classes of models. For example, the search for new resonance relies on the assumption that the new physics appears as the exchange of a single new particle.  In the following, we will rather assume that the new physics is too heavy to be produced directly by the experiments but rather shows up as new interactions between the known particles. Therefore, those effects can be described by an effective Lagrangian~\cite{Weinberg:1978kz,Weinberg:1979pi}.
Effective field theories (EFT) will be introduced in section~\ref{sec:eft} and applied to electroweak gauge bosons productions in sections~\ref{sec:ppWW} and \ref{sec:scat}.

\section{Effective Field Theories}\label{sec:eft}

An effective Lagrangian is built from the low energy degrees of freedom only, the SM fields including the Higgs boson in our case. As supported by the data, the operators should also preserve the $SU(3)_c\otimes SU(2)_L \otimes U(1)_Y$ symmetries as well as the baryon and lepton symmetries. Only even dimension operators can be built with those conditions~\cite{Degrande:2012wf}. All the operators with dimension equal to two or four constitute the SM Lagrangian\footnote{The only missing operator induces CP violation by the strong interaction.} while the higher dimension operators are suppressed by the new scale and induce new interactions,
\begin{equation}
\mathcal{L}=\mathcal{L}_{SM}+\sum_{d=6}^\infty \sum_{i} \frac{c_i}{\Lambda^{d-4}} \mathcal{O}^d_i,\label{eq:lag}
\end{equation}
where d is the dimension of the operators $\mathcal{O}^d_i$, $\Lambda$ is the new scale and $c_i$ are coefficients that can be derived from the complete theory including the heavy degrees of freedom. Far from the new physics scale, the Lagrangian \eqref{eq:lag} reduces to the SM one as required. As the energy probed gets closer to the new physics scale, interactions from the dimension-six operators become relevant. However, the theory is predictive even if their coefficients are kept as free parameters to be model independent because this set of operators is finite. Finally, all the operators have large effects at or above the new physics scale and the effective theory is no longer useful. Consequently, the EFT is only valid below the new scale.\\
59 dimension-six operators can be built out of the SM fields for only one generation of fermion~\cite{Buchmuller:1985jz,Grzadkowski:2003tf,Fox:2007in,AguilarSaavedra:2008zc,AguilarSaavedra:2009mx,Grzadkowski:2010es}. Despite that this number increases significantly if three flavors are considered, the effective extension of the SM is predictive and only a few operators contribute to a particular process~\cite{Leung:1984ni,Buchmuller:1985jz}. Usually, the different operators can be distinguished because they do not affect the same observables and/or  contribute to several processes as it will be illustrated in sections \ref{sec:effects} and \ref{sec:scat}. As a matter of fact, most of the operators induce several vertices and contribute to more than one amplitude due to gauge symmetries. 
Gauge symmetries also cause EFT to be more predictive than the alternative anomalous couplings approach~\cite{Gaemers:1978hg}. In fact, anomalous couplings are based only on Lorentz invariance. While electromagnetic gauge invariance is often imposed afterwards to reduce the number of parameters, the $SU(2)_L$ symmetry is ignored. 
As there is no high scale in the anomalous couplings approach, there are no theoretical motivations to truncate the series of operators. Additionally, the Lagrangian is expected to be valid at any scale. However, unitarity is broken by the new interactions at some scale and requires the introduction of arbitrary form factors~\cite{Zeppenfeld:1987ip,Baur:1987mt,Hagiwara:1989mx}. On the contrary, unitarity is violated only at or above the new scale in an EFT and no form factors are needed in its validity region.
Finally, the last important virtue of EFT is to be renormalizable in the modern sense, \textit{i.e.} order by order in $\Lambda$. Consequently, effective Lagrangians can be used to compute loops (see for example~\cite{Arzt:1992wz}) and the operators can be constrained both by direct and indirect measurements. 

\section{EFT for $pp\to WW$}\label{sec:ppWW}

\subsection{The operators}
\label{sec:op}

We focus here on the operators that affect the interactions between the electroweak bosons and assume that those modifying the interaction between the quarks and the vector bosons are constrained by other processes like Drell-Yan or the Z decays. Three CP-conserving dimension-six operators,
\begin{eqnarray}
{\cal O}_{WWW}&=&\mbox{Tr}[W_{\mu\nu}W^{\nu\rho}W_{\rho}^{\mu}]\nonumber\\
{\cal O}_W&=&(D_\mu\Phi)^\dagger W^{\mu\nu}(D_\nu\Phi)\label{eq:OW}\nonumber\\
{\cal O}_B&=&(D_\mu\Phi)^\dagger B^{\mu\nu}(D_\nu\Phi),
\end{eqnarray}
and two CP violating dimension-six operators,
\begin{eqnarray}
{\cal O}_{\tilde WWW}&=&\mbox{Tr}[{\tilde W}_{\mu\nu}W^{\nu\rho}W_{\rho}^{\mu}]\nonumber\\
{\cal O}_{\tilde W}&=&(D_\mu\Phi)^\dagger {\tilde W}^{\mu\nu}(D_\nu\Phi),
\end{eqnarray}
satisfy this requirement. In the operators definition, $\Phi$ is the Higgs doublet and the covariant derivative and the strength field tensors are defined by
\begin{align}
D_\mu & = \partial_\mu + \frac{i}{2} g \tau^I W^I_\mu + \frac{i}{2} g' B_\mu \nonumber\\
W_{\mu\nu} & = \frac{i}{2} g\tau^I (\partial_\mu W^I_\nu - \partial_\nu W^I_\mu
	+ g \epsilon_{IJK} W^J_\mu W^K_\nu )\nonumber\\
B_{\mu \nu} & = \frac{i}{2} g' (\partial_\mu B_\nu - \partial_\nu B_\mu).
\end{align}
Other basis of operators can be chosen. However, the triple gauge coupling and the weak boson masses are influenced by different operators at tree-level in this basis~\cite{Grojean:2006nn}. Consequently, the coefficients of those operators do not have strong constraints from electroweak precision tests. 

\subsection{Comparison to anomalous couplings}\label{sec:ano}

The anomalous couplings Lagrangian is \cite{Hagiwara:1986vm}
\begin{eqnarray}
\!\!\!\!{\mathcal L_{AC}}&=&ig_{WWV}\left(g_1^V(W_{\mu\nu}^+W^{-\mu}-W^{+\mu}W_{\mu\nu}^-)V^\nu \right.\nonumber\\
&&+\kappa_VW_\mu^+W_\nu^-V^{\mu\nu}+\frac{\lambda_V}{M_W^2}W_\mu^{\nu+}W_\nu^{-\rho}V_\rho^{\mu}\nonumber\\
&&+ig_4^VW_\mu^+W^-_\nu(\partial^\mu V^\nu+\partial^\nu V^\mu)
\nonumber\\&&\left.
-ig_5^V\epsilon^{\mu\nu\rho\sigma}(W_\mu^+\partial_\rho W^-_\nu-\partial_\rho W_\mu^+W^-_\nu)V_\sigma
\right.\nonumber\\&&\left.
+\tilde{\kappa}_VW_\mu^+W_\nu^-\tilde{V}^{\mu\nu}
+\frac{\tilde{\lambda}_V}{m_W^2}W_\mu^{\nu+}W_\nu^{-\rho}\tilde{V}_\rho^{\mu}
\right)
\label{eq:Lac}
\end{eqnarray}
where $V=\gamma,Z$;  $W_{\mu\nu}^\pm = \partial_\mu W_\nu^\pm - \partial_\nu W_\mu^\pm$, $V_{\mu\nu} = \partial_\mu V_\nu - \partial_\nu V_\mu$, and the overall coupling constants are defined as $g_{WW\gamma}=-e$ and $g_{WWZ}=-e\cot\theta_W$. This Lagrangian contains 14 free parameters. Yet, nothing forbits the addition of further terms with extra derivatives. Electromagnetic gauge invariance implies $g_1^\gamma =1$ and $g_4^\gamma=g_5^\gamma = 0$. Ultimately, the triple anomalous couplings for the charged gauge bosons are described using five C- and P-conserving parameters $g_1^Z,\, \kappa_\gamma,\, \kappa_Z,\, \lambda_\gamma$ and $ \lambda_Z$ and six C- and/or P-violating parameters $g_4^Z,\, g_5^Z,\, \tilde{\kappa}_\gamma,\, \tilde{\kappa_Z},\, \tilde{\lambda}_\gamma$ and $ \tilde{\lambda_Z}$. This approach has six extra parameters compared to the EFT. However, the anomalous couplings can be derived from the EFT,
\begin{eqnarray}
g_1^Z & = & 1+c_W\frac{m_Z^2}{2\Lambda^2}\nonumber\\
\kappa_\gamma & = & 1+(c_W+c_B)\frac{m_W^2}{2\Lambda^2}\nonumber\\
\kappa_Z & = & 1+(c_W-c_B\tan^2\theta_W)\frac{m_W^2}{2\Lambda^2}\nonumber\\
\lambda_\gamma & = & \lambda_Z = c_{WWW}\frac{3g^2m_W^2}{2\Lambda^2}\nonumber\\
g_4^V &=& g_5^V=0\nonumber\\
\tilde{\kappa}_\gamma & = &c_{\tilde{W}}\frac{m_W^2}{2\Lambda^2}\nonumber\\
\tilde{\kappa}_Z & = &-c_{\tilde{W}}\tan^2\theta_W\frac{m_W^2}{2\Lambda^2}\nonumber\\
\tilde{\lambda}_\gamma & = & \tilde{\lambda}_Z = c_{\tilde{W}WW}\frac{3g^2m_W^2}{2\Lambda^2}.\label{eq:fromEFT}
\end{eqnarray}
The above expressions for anomalous coupling are constant,  \textit{i.e.} they do not depend on the momenta of the vector boson. 
Alternatively, EFT predictions require two relations \cite{Hagiwara:1993ck},
\begin{eqnarray}
\Delta g_1^Z&=&\Delta \kappa_Z + \tan^2\theta_W \Delta \kappa_\gamma\nonumber\\
\lambda_\gamma &=& \lambda_Z,\label{eq:relCP}
\end{eqnarray}
where $\Delta g_1^Z = g_1^Z - 1$ and $\Delta \kappa_{\gamma,Z} = \kappa_{\gamma,Z} - 1$ for the CP-conserving couplings
and four relations,
\begin{eqnarray}
0&=&\tilde \kappa_Z + \tan^2\theta_W \tilde \kappa_\gamma\nonumber\\
\tilde\lambda_\gamma &=& \tilde\lambda_Z\nonumber\\
g_4^Z&=&g_5^Z=0,\label{eq:relCPV}
\end{eqnarray}
for the C- and/or P-violating couplings. 

\subsection{The PDG constraints}
\label{sec:const}

The PDG~\cite{Beringer:1900zz} constraints on triple gauge couplings,
\begin{eqnarray}
g_1^Z&=&0.984^{+0.022}_{-0.019}\nonumber\\
\kappa_\gamma&=&0.979^{+0.044}_{-0.045}\nonumber\\
\lambda_\gamma&=&0.028^{+0.020}_{-0.021}\nonumber\\
\tilde\kappa_\gamma&=&0.12^{+0.06}_{-0.04}\nonumber\\
\tilde\lambda_\gamma&=&0.09\pm0.07,
\end{eqnarray}
 are obtained from LEP mearurements of one parameter at a time. Although the notations from anomalous couplings is used, the measurements have been done imposing the relations from EFT of Eqs. \eqref{eq:relCP} and \eqref{eq:relCPV}. Inverting the system in Eq.~\eqref{eq:fromEFT} and ignoring the correlations, we obtain the following constraints on the coefficients of the dimension-six operators,
\begin{eqnarray}
c_{WWW}/\Lambda^2&\in&\left[-11.9,1.94\right]\text{TeV}^{-2}\nonumber\\
c_{W}/\Lambda^2&\in&\left[-8.42,1.44\right]\text{TeV}^{-2}\nonumber\\
c_{B}/\Lambda^2&\in&\left[-7.9,14.9\right]\text{TeV}^{-2}\nonumber\\
c_{\tilde WWW}/\Lambda^2&\in&\left[-185.3,-82.4\right]\text{TeV}^{-2}\nonumber\\
c_{\tilde W}/\Lambda^2&\in&\left[-39.3,-4.9\right]\text{TeV}^{-2}
\end{eqnarray}
at 68\% C.L..
Although Tevatron measurements of the anomalous couplings are present in the PDG, they are not included in the combination. Tevatron experiments do not always use the relations in  Eqs. \eqref{eq:relCP} and \eqref{eq:relCPV}. Moreover, they use form factors most of the time such that the constraints depend on the center of mass energy. Consequently, a combination of the results of all the accelerators is almost impossible for anomalous couplings. On the contrary, the EFT allows a combination of the constraints from various machines.


\subsection{Unitarity bound}
\label{sec:uni}

\begin{figure}[h]
\centering
\includegraphics[width=0.49\textwidth,clip]{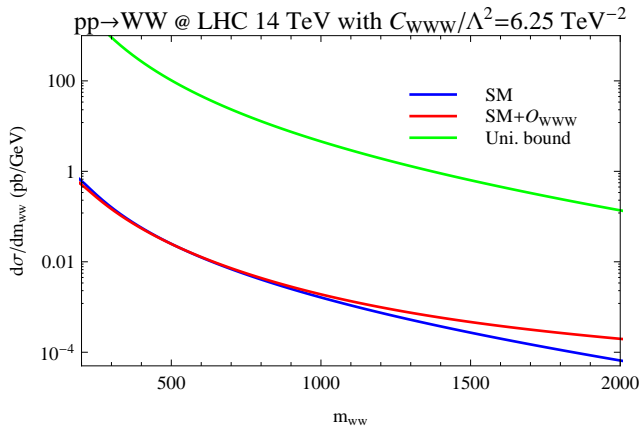}
\caption{Invariant mass distributions for the production of two W bosons for the SM in blue, for the SM and $\mathcal{O}_{WWW}$ in red and for the unitarity bound in green.}
\label{fig:UOwww}       
\end{figure}

At large invariant mass, the SM cross-section falls in $1/s$ like the unitarity bound such that they never cross. Dimension-six operators as well as anomalous couplings in Eq.~\eqref{eq:Lac} do not have this behavior. Their contributions are most of the time either constant for their interferences with the SM or grow linearly with $s$ for their squared amplitudes as illustrated on Fig.~\ref{fig:UOwww}. 
Eventually, the new physics contribution will reach the unitarity bound proving that the two Lagrangians introduced above cannot be used to arbitrary high energies. Contrary to EFT, no such assumption is made for anomalous couplings. The unitarity bound is then avoided by adding form factors to the anomalous couplings, \textit{i.e.} by multiplying them by a function of the invariant mass such that their contributions vanish at large $s$. While the experimental bounds depend on which function is chosen for the form factors, the function is arbitrary.
However, this unitarity bound is about two orders of magnitude above the prediction from EFT using the constraints from LEP as shown on Fig.~\ref{fig:UOwww}. Consequently, the form factors are not needed to describe the data.

\subsection{The effects}
\label{sec:effects}

\begin{figure}[h]
\centering
\includegraphics[width=0.49\textwidth,clip]{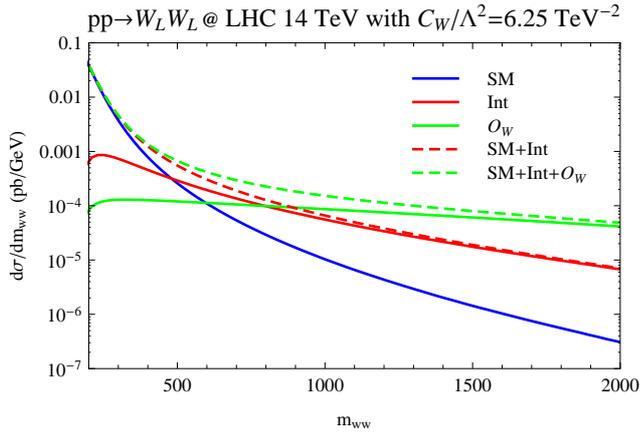}
\caption{Invariant mass distributions for the production of two longitudinally polarized W bosons for the SM (Solid blue line), for the interference between the SM and $\mathcal{O}_W$ (Solid red line), for $\mathcal{O}_W$ squared amplitude (Solid green line), for the sum of the SM and its interference with the dimension-six operator (Dashed red line) and for the sum of all the contributions (Dashed green line).}
\label{fig:OwLL}       
\end{figure}

\begin{figure}[h]
\centering
\includegraphics[width=0.49\textwidth,clip]{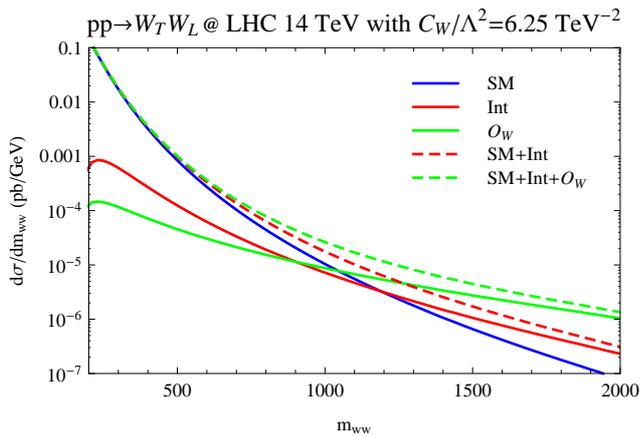}
\caption{Invariant mass distributions for the production of one longitudinally polarized W boson and one transversely polarized W boson with the same color convention as in Fig.~\ref{fig:OwLL}.}
\label{fig:OwLT}      
\end{figure}

\begin{figure}[h]
\centering
\includegraphics[width=0.49\textwidth,clip]{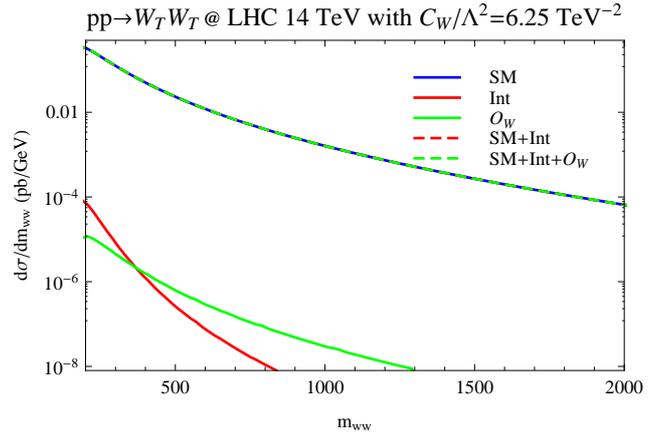}
\caption{Invariant mass distributions for the production of two transversely polarized W bosons with the same color convention as in Fig.~\ref{fig:OwLL}.}
\label{fig:OwTT}      
\end{figure}

The behavior mentioned in section~\ref{sec:uni} can be used to distinguishes the new physics contributions from the SM one but does not allow to differentiate them. On the contrary, the contributions for specific helicities of the W bosons are different  for each of the CP-even operators as illustrated by Tab.~\ref{tab:highs} .
\begin{table}[h]
\centering
\begin{tabular}{|c|c|c|c|c|}
\hline
  & SM & $\mathcal{O}_{WWW}$& $\mathcal{O}_{W}$& $\mathcal{O}_{B}$  \\\hline
$W_LW_L$ & $1/s$ & $0$ &$1(s)$&$1(s)$ \\\hline
$W_LW_T$ & $1/s^2$ & $1/s(1)$ &$1/s(1)$&$1/s(1)$ \\\hline
$W_TW_T$ & $1/s$ & $1/s(s)$ &$1/s^2(1/s)$&$0$ \\\hline
\end{tabular}
\caption{Behavior at large $s$ of the SM, its interferences with the dimension-six operators and,  in parentheses, the squared amplitudes of the dimension-six operators for the different polarizations of the two W bosons.}
\label{tab:highs}     
\end{table}
As a matter of fact, all the interferences are constant except for $\mathcal{O}_{WWW}$\footnote{The exchange of a triplet is suppressed in the SM and is responsible for the suppression of the interference.}  and the squared new physics amplitudes grow like $s$  at large invariant mass after summing over the W polarizations. However, $\mathcal{O}_{WWW}$ contributions to a pair of longitudinally or transversally polarized W bosons differ from those of the two other CP-even operators.  $\mathcal{O}_B$ can in principle be distinguished form $\mathcal{O}_W$ by their contributions to two transversally polarized W bosons. The extra $s$ factors in the cross-sections of those two operators are coming from the longitudinal polarization vectors and not from their vertices because they contain only one momentum like the SM and unlike $\mathcal{O}_{WWW}$.  Therefore their largest contributions are to $pp\to W_LW_L$ while the smallest one is to $pp\to W_T W_T$ (see Figs.~\ref{fig:OwLL}, \ref{fig:OwLT} and \ref{fig:OwTT}). Consequently, their difference is challenging from the experimental point of view. 
Moreover, the largest SM contribution is precisely to $pp\to W_T W_T$ such that the new physics is invisible in this channel as shown on Fig.~\ref{fig:OwTT}. 
The opposite orders of the contributions to the various W polarizations between the SM and the two dimension-six operators explain the suppression of the total interferences. \\
Finally, the invariant mass distributions show explicitly that EFT are valid only at low energies and an estimate of the scale at which the expansion in $1/\Lambda^2$ breaks can be extracted from the intersections of the solid line in Fig.~\ref{fig:OwLL}. Around 600 GeV, the order $\Lambda^0$ (SM), $\Lambda^{-2}$ (interference) and $\Lambda^{-4}$ (operator squared amplitude) terms become all of the same size. This estimate is consistent with the scale obtain if $c_{W}=1$, \textit{i.e.} $\Lambda=400$ GeV. While Fig.~\ref{fig:OwLT} can also be used, Fig.~\ref{fig:OwTT} does not provide a reliable estimate as the new physics contributions are suppressed.

\section{EFT for $pp\to WWW$,  $pp\to WWjj$}\label{sec:scat}

The similarity of those two processes is that they both probe the four W amplitude. Ignoring again the modification of the quarks electroweak interactions, new physics contributions arise through the modification of the triple and quartic gauge boson vertices. While the two sets of vertices depend on different parameters in the anomalous couplings approach, they are induced by the same dimension-six operators. In fact, gauge invariance requires relations between the triple and the quartic gauge boson vertices. Those relations are guaranteed in EFT by construction. On the contrary, they need to be computed from the Ward identities for massive vectors for the anomalous couplings.\\
Since the largest new physics contributions to all the electroweak vector self couplings arise from the same dimension-six operators,  $pp\to WWW$ and  $pp\to WWjj$ can be used to further constrain their coefficients. However, some dimension-eight operators affect only the quartic gauge coupling \cite{Eboli:2006wa} and suggest to check the consistency between the measurements in diboson and triboson productions or boson scattering before combination. 

\section{Conclusion}

Only three CP-even and two CP-odd operators affect the productions of electroweak gauge bosons. As a consequence, the effective extension of the SM is more predictive than the anomalous couplings approach. The EFT is also simpler because gauge invariance is guaranteed by construction and no form factors are needed. This last virtue appears to be quite important to combine the results of various accelerators. Additionally, the direct and indirect constraints can be combined because the theory is renormalizable in the modern sense.
Recently, tools to automate the introduction of EFT in events generators~\cite{Degrande:2011ua} and to handle the new structures of their vertices~\cite{deAquino:2011ub}  have been developed. In particular, the EFT for the productions of weak vector bosons is available~\cite{EWdim6}.

\bibliography{biblio}

\begin{thebibliography}{23}

\bibitem{Weinberg:1978kz}
S.~Weinberg, Physica \textbf{A96}, 327 (1979)

\bibitem{Weinberg:1979pi}
S.~Weinberg, Rev.Mod.Phys. \textbf{52}, 515 (1980)

\bibitem{Degrande:2012wf}
C.~Degrande, N.~Greiner, W.~Kilian, O.~Mattelaer, H.~Mebane et~al. (2012),
  \texttt{1205.4231}

\bibitem{Buchmuller:1985jz}
W.~Buchmuller, D.~Wyler, Nucl.Phys. \textbf{B268}, 621 (1986)

\bibitem{Grzadkowski:2003tf}
B.~Grzadkowski, Z.~Hioki, K.~Ohkuma, J.~Wudka, Nucl.Phys. \textbf{B689}, 108
  (2004), \texttt{hep-ph/0310159}

\bibitem{Fox:2007in}
P.J. Fox, Z.~Ligeti, M.~Papucci, G.~Perez, M.D. Schwartz, Phys.Rev.
  \textbf{D78}, 054008 (2008), \texttt{0704.1482}

\bibitem{AguilarSaavedra:2008zc}
J.~Aguilar-Saavedra, Nucl.Phys. \textbf{B812}, 181 (2009), \texttt{0811.3842}

\bibitem{AguilarSaavedra:2009mx}
J.~Aguilar-Saavedra, Nucl.Phys. \textbf{B821}, 215 (2009), \texttt{0904.2387}

\bibitem{Grzadkowski:2010es}
B.~Grzadkowski, M.~Iskrzynski, M.~Misiak, J.~Rosiek, JHEP \textbf{1010}, 085
  (2010), and ref. [10-13] therein, \texttt{1008.4884}

\bibitem{Leung:1984ni}
C.N. Leung, S.~Love, S.~Rao, Z.Phys. \textbf{C31}, 433 (1986)

\bibitem{Gaemers:1978hg}
K.~Gaemers, G.~Gounaris, Z.Phys. \textbf{C1}, 259 (1979)

\bibitem{Zeppenfeld:1987ip}
D.~Zeppenfeld, S.~Willenbrock, Phys.Rev. \textbf{D37}, 1775 (1988)

\bibitem{Baur:1987mt}
U.~Baur, D.~Zeppenfeld, Phys.Lett. \textbf{B201}, 383 (1988)

\bibitem{Hagiwara:1989mx}
K.~Hagiwara, J.~Woodside, D.~Zeppenfeld, Phys.Rev. \textbf{D41}, 2113 (1990)

\bibitem{Arzt:1992wz}
C.~Arzt, M.~Einhorn, J.~Wudka, Phys.Rev. \textbf{D49}, 1370 (1994),
  \texttt{hep-ph/9304206}

\bibitem{Grojean:2006nn}
C.~Grojean, W.~Skiba, J.~Terning, Phys.Rev. \textbf{D73}, 075008 (2006),
  \texttt{hep-ph/0602154}

\bibitem{Hagiwara:1986vm}
K.~Hagiwara, R.~Peccei, D.~Zeppenfeld, K.~Hikasa, Nucl.Phys. \textbf{B282}, 253
  (1987)

\bibitem{Hagiwara:1993ck}
K.~Hagiwara, S.~Ishihara, R.~Szalapski, D.~Zeppenfeld, Phys.Rev. \textbf{D48},
  2182 (1993)

\bibitem{Beringer:1900zz}
J.~Beringer et~al. (Particle Data Group), Phys.Rev. \textbf{D86}, 010001 (2012)

\bibitem{Eboli:2006wa}
O.~Eboli, M.~Gonzalez-Garcia, J.~Mizukoshi, Phys.Rev. \textbf{D74}, 073005
  (2006), \texttt{hep-ph/0606118}

\bibitem{Degrande:2011ua}
C.~Degrande, C.~Duhr, B.~Fuks, D.~Grellscheid, O.~Mattelaer et~al.,
  Comput.Phys.Commun. \textbf{183}, 1201 (2012), \texttt{1108.2040}

\bibitem{deAquino:2011ub}
P.~de~Aquino, W.~Link, F.~Maltoni, O.~Mattelaer, T.~Stelzer,
  Comput.Phys.Commun. \textbf{183}, 2254 (2012), \texttt{1108.2041}

\bibitem{EWdim6}
\texttt{https://cp3.irmp.ucl.ac.be/projects/\\madgraph/wiki/Models/EWdim6}

\end{thebibliography}

\end{document}